\documentclass{article}
\newcommand{\nn}{\nonumber\\}
\newcommand{\p}[1]{(\ref{#1})}

\newcommand{\bu}{{\bar u}}
\newcommand{\bl}{{\bar \lambda}}
\newcommand{\bL}{{\overline\Lambda}}

\newcommand{\bD}{{\overline D}{}}
\newcommand{\bnab}{{\overline\nabla }{}}
\newcommand{\cF}{{\cal F}}

\newcommand{\cN}{{\cal N}}

\newcommand{\cQ}{{\cal Q}}

\newcommand{\cbF}{\overline{\cal F}}

\newcommand{\ba}{\begin{array}}
\newcommand{\ea}{\end{array}}
\newcommand{\be}{\begin{equation}}
\newcommand{\ee}{\end{equation}}
\newcommand{\bea}{\begin{eqnarray}}
\newcommand{\eea}{\end{eqnarray}}
\newcommand{\bi}{\begin{itemize}}
\newcommand{\ei}{\end{itemize}}

\newcommand{\bbib}[1]{\begin{thebibliography}{#1}}
\newcommand{\ebib}{\end{thebibliography}}

\usepackage{amscd,amsmath,amssymb}
\begin{document}
\thispagestyle{empty}
\vspace{2cm}
\begin{flushright}
%Version on \today \\[3cm]
\end{flushright}

\begin{center}
{~}\\
\vspace{3cm}
{\Large\bf N=8 supersymmetric mechanics on the sphere $S^3$}\\
\vspace{2cm}
{\large \bf S.~Bellucci${}^{a}$, S.~Krivonos${}^{b}$ and A.~Sutulin${}^{b}$ }\\
\vspace{2cm}
{\it ${}^a$INFN-Laboratori Nazionali di Frascati, Via E. Fermi 40,
00044 Frascati, Italy}\\
{\tt bellucci@lnf.infn.it} \\\vspace{0.5cm}
{\it ${}^b$ Bogoliubov  Laboratory of Theoretical Physics, JINR,
141980 Dubna, Russia}\\
{\tt krivonos, sutulin@theor.jinr.ru} \\ \vspace{2.5cm}
\end{center}

\begin{center}
{\bf Abstract}
\end{center}
Starting from quaternionic $N=8$ supersymmetric mechanics we perform a reduction over a
bosonic radial variable, ending up with a nonlinear off-shell supermultiplet with three bosonic end eight fermionic
physical degrees of freedom. The geometry of the bosonic sector of the most general sigma-model type action is
described by an arbitrary function obeying the three dimensional Laplace equation on the sphere $S^3$.
Among the bosonic components of this new supermultiplet there is a constant which gives rise to potential terms.
After dualization of this constant one may come back to the supermultiplet with four physical bosons.
However, this new supermultiplet is highly nonlinear. The geometry of the corresponding sigma-model action
is briefly discussed.

\hfil
\newpage
\section{Introduction}
Despite the many common structures the mechanics with extended supersymmetries shares with its higher
dimensional counterparts (see e.g. \cite{intro}) it also possesses some  rather specific features which
cannot be even imagined in other dimensions. Among the most impressive examples one may find the one dimensional
$N=4,8$ supermultiplets without auxiliary components or even without physical bosons \cite{ex} as well as
plenty of off-shell nonlinear supermultiplets. Just a new $N=8$ nonlinear supermultiplet and its
action are the subject of the present Letter.

The idea about the possibility of the existence of the nonlinear supermultiplets in $N=4, d=1$ supersymmetric theories has been
formulated for the first time in \cite{IL}. In a short time, some of the nonlinear $N=4$ supermultiplets
have been explicitly constructed in \cite{nlin}. Later on, it has been understood that almost all
$N=4$ supermultiplets, linear and nonlinear ones can be obtained proceeding from the ``root'' supermultiplet
and dualizing the physical bosons  into auxiliary components \cite{root}. Alternatively, the superfield
procedure for constructing linear and nonlinear supermultiplets starting from the root one
has been recently proposed in \cite{ID}. Finally, after some preliminary consideration \cite{DK}
it was found that there are infinitely many $N=4$ nonlinear supermultiplets with a functional
freedom in the definition \cite{ks}.

When passing to $N=8, d=1$ supersymmetry, the situation becomes more complicated. Until now only two
nonlinear off shell $N=8$ supermultiplets are known - those with two \cite{8nlin} and four \cite{mar,EI}
physical bosons were constructed off-shell. Other known nonlinear supermultiplets, for example
those discussed in \cite{dual1}, have been constructed with the help of duality transformations and
had been fully described only on-shell. One may wonder why we need nonlinear supermultiplets and
which additional problems appear in the case of $N=8$ supersymmetric mechanics. The answer to these
questions
is that the $N=8$ supersymmetry puts a rather strong restriction on the metric of the sigma-model
part of the action -
for all linear supermultiplets the metric has to be conformally flat, with the conformal factor
obeying a proper Laplace equation. Thus, the hyper-K\"ahler metrics never show up within the
models with linear supermultiplets. Moreover, the reduction procedure which perfectly worked in the $N=4$
case is not so useful for the $N=8$ case. The explanation is very simple - in order to dualize the physical
boson into the auxiliary component one should choose a metric which does not contain this boson. Only with
such a condition it is possible to turn this physical boson into the auxiliary component. But in almost
all cases this cannot be achieved, because the metrics is subjected to the Laplace equation.
Nevertheless, there is one special case where the reduction can be performed. It corresponds
to the $N=8$ supersymmetric mechanics with four physical bosonic components. Only in this case the
solution of the four dimensional Laplace equation admits the reduction over radii and results
in a three dimensional system with $N=8$ supersymmetry. Just this case will be analyzed in this Letter.
We start from the known quaternionic four dimensional $N=8$ mechanics \cite{antonio} and
perform the reduction over the ``radii'' variable. In such a way we will get the $N=8$ supersymmetric
system on the sphere $S^3$ (Section 2). Then we will write the proper superfield constraint, which
follows from the constraints on the four dimensional supermultiplet upon reduction, and construct
the most general action (Section 3). In Section 4 we consider the potential terms and
perform another dualizations of the coupling constant which pushes us back to a new nonlinear
four dimensional supermultiplet. For this supermultiplet we find that the geometry of the bosonic sigma-model
is neither conformally flat nor of a hyper-K\"ahler type. In fact, this geometry is just the one
which was previously found in the $(4,0)$ heterotic sigma models in $d=2$ \cite{heterotic}.

\section{N=8 supersymmetric mechanics on the sphere $S^3$}

As we already mentioned in the Introduction, our idea is to perform the reduction over radii to pass from
quaternionic four dimensional $N=8$ mechanics \cite{antonio} to three dimensional mechanics with the
sphere $S^3$ in the bosonic part of the action.

The supermultiplet which has been used to construct the quaternionic mechanics is described by a
quartet of $N=8$ superfields $\cQ^{ia}$ depending on the coordinates of the
$N=8$, $d=1$ superspace $\mathbb{R}^{(1|8)}$. These superfields are subjected to the following constraints
\cite{bikl2}:

\be
D^{(i}_A \cQ^{j)a}   = 0\,, \qquad
 \nabla^{(a}_{\alpha} \cQ^{ib)}  = 0\,.
\label{conQ}
\ee
Here $i,\,a,\,A,\,\alpha=1,\,2$ are doublet indices of four $SU(2)$ subgroups of
the automorphism group of $N=8$ superspace, and the spinor covariant derivatives are defined to obey
the algebra
\be
\left\{ D^{iA}, D^{jB} \right\} = 2i \epsilon^{ij}\epsilon^{AB}\partial_t, \quad
\left\{ \nabla^{a\alpha}, \nabla^{b\beta} \right\} = 2i \epsilon^{ab}\epsilon^{\alpha\beta}\partial_t.
\ee

The component {\it on-shell} form of the corresponding action with arbitrary metric function $G$
depending on four physical bosonic fields is given by
\bea
\label{actionQ}
S&=& \int dt\, \left[ G \left( \dot{q}^{ia}\dot{q}_{ia}
+ \frac{i}{2}\xi^{i\alpha}\dot{\xi}_{i\alpha}
+ \frac{i}{2}\psi^{aA}\dot{\psi}_{aA}\right)+
\frac{i}{2} \frac{\partial G}{\partial q^{ia}} \left(  \xi^{i\alpha}\xi_\alpha^k\; \dot{q}^a_k
+ \psi^{aA}\psi_A^b\; \dot{q}^i_b\right) \right. + \nn
&& \left. \frac{1}{8} \left( \frac{\partial^2 G}{\partial q^{ia} \partial q^{kb}}
- 2 G^{-1}\,\frac{\partial G}{\partial q^{ia}} \frac{\partial G}{\partial q^{kb}}\right)\,
 \xi^{i\alpha}\xi_\alpha^k\;\psi^{aA}\psi_A^b\right] \;.
\eea
The invariance under  $N=8$ supersymmetry imposes the additional constraint on the metric $G$ to be
harmonic
\be \frac{\partial^2}{\partial q^{ia} \partial q_{ia}} G=0.
\ee
It is easy to see that in the special case of the metric function
\be
G = \frac{1}{{q}^{ia}\,{q}_{ia}}\,,
\label{metricQ}
\ee
the four-fermion term in \p{actionQ} is canceled.
Then, introducing new bosonic and fermionic variables
\bea
&&
{q}^{ia} = \tilde {q}^{ia} e^{u/2}\,, \quad
\tilde {q}^{ia}\tilde {q}^k_a = \epsilon^{ki}\,, \quad
\tilde {q}^{ia} \tilde {q}_i^b = \epsilon^{ba}\,,
\nn
&&
\tilde \psi^{iA}  = e^{-u/2} \tilde {q}^i_a \psi^{aA}\,, \quad
\tilde \xi^{a \alpha} =  e^{-u/2} \tilde {q}_i^a {\xi}^{i\alpha}\,,
\label{newferm}
\eea
one finds that the action (\ref{actionQ}) is reduced to the sum of a bosonic sigma model
type action and the action for free fermions
\be
S = \frac{1}{2}\,\int dt\, \Big \{\frac{1}{2}\,
(\dot u)^2 + \dot{\tilde q}^{ia}\, \dot{\tilde q}_{ia}
+ \frac{i}{2}\,
\tilde \xi^{a \alpha}\, \dot{\tilde \xi}_{a \alpha}
+ \frac{i}{2}\, \tilde \psi^{iA}\, \dot{\tilde \psi}_{iA}
\Big \}
\ee
If we replace the time derivative of the field $u$ by an auxiliary field $A=\dot u$ and then exclude $A$ by
its equation of motion, we will get just $N=8$ mechanics on the sphere $S^3$.
What is really interesting is that the $N=8$ supersymmetrization of this $S^3$ is achieved by
adding eight free fermions. Let us remind, that just the same phenomenon appears in the case of the $N=4$
supersymmetrization of the spheres $S^2$ \cite{nlin} and $S^3$ \cite{nlin1,nlin2}.

\setcounter{equation}0
\section{Three dimensional mechanics: superfields and components}

The example of straightforward reduction presented in the previous Section gives only the simplest variant
of the three dimensional action. In order to construct the most general sigma-model action one should
perform the reduction in terms of superfields. Fortunately, this is rather easy to do. Indeed,
let us introduce new $N=8$ quartet superfields $\cN^{ia}$ as follows:
\be
\cN^{ia} = \frac{\cQ^{ia}}{|\cQ|}\,, \quad |\cQ|^2 = \cQ^{ia} \cQ_{ia} \,.
\label{NF}
\ee
Clearly, these new defined superfields $\cN^{ia}$ do not include the ``radii component'' $|\cQ|^2$ from
the supermultiplet $\cQ^{ia}$. Moreover, in virtue of  (\ref{conQ}), the superfields $\cN^{ia}$  obey
the closed set of constraints
\be\label{basiccon}
\cN_a^{\,(k} D^i_A \cN^{j) a} = 0\,, \qquad
\cN_i^{\,(a} \nabla^b_{\alpha} \cN^{c)i} = 0, \qquad N^{ia}N_{ia}=2\,.
\ee
Thus, the structure of our nonlinear supermultiplet is completely defined by the constraints \p{basiccon}.

The concise form of the constraints \p{basiccon} is not practically useful. It is convenient to use
instead the following parametrization for them (which solves the last, algebraic constraint in \p{basiccon}):
\be\label{u}
\cN^{\,ia} = \Big (\cN^{\,11}\,,\cN^{\,12}\,,\cN^{\,21}\,,\cN^{\,22} \Big )
= \Big ( \frac{1}{\sqrt{1+u\bu}}\,u\,,\;\;
\frac{e^{i/2 \phi}}{\sqrt{1+u\bu}}\,,\;\;
- \frac{e^{-i/2 \phi}}{\sqrt{1+u\bu}}\,,\;\;
\frac{1}{\sqrt{1+u\bu}}\,\bu\, \Big )\,.
\ee
In this parametrization the constraints \p{basiccon} read
\bea\label{ucon}
D^i \left( e^{-\frac{i}{2}\phi}u\right) =0,\;\bD_i \left( e^{\frac{i}{2}{\phi}}\bu\right) =0,& \quad &
\nabla^a \left( e^{\frac{i}{2}{\phi}}u\right) =0,\;\bnab_a \left( e^{-\frac{i}{2}{\phi}}\bu\right) =0,\nn
D^i \left( e^{-\frac{i}{2}{\phi}}\bu\right) =-i\bD^i \phi,\;\bD_i \left( e^{\frac{i}{2}{\phi}}u\right) =-iD_i \phi,& \quad &
\nabla^a \left( e^{\frac{i}{2}{\phi}}\bu\right) =-i\bnab^a \phi,\;\bnab_a \left( e^{-\frac{i}{2}{\phi}}u\right) =-i\nabla_a \phi.
\eea
Here, we redefine the covariant derivatives as
$$D^i=\{D^{11},D^{12}\},\; \bD^i =\{D^{21},D^{22}\},\quad \nabla^a=\{\nabla^{11},\nabla^{12}\},\; \bnab^a=\{\bnab^{21},\bnab^{22}\} ,$$
which now anticommute as follows:
\be\label{sderiv}
\left\{ D^{i},\bD_{j}\right\}=-2i \delta^{i}_j \partial_t, \qquad
\left\{ \nabla^{a},\bnab_{b}\right\}=-2i \delta^{a}_{b}\partial_t.
\ee

Besides the superfields $\left\{ \phi, u, \bu\ \right\}$  it is rather convenient to introduce the related sets
of superfields $\left\{ \phi, \Lambda, \bL\ \right\}$ and $\left\{ \phi, \lambda, \bl\ \right\}$ as
\be\label{lambda}
e^{\frac{i}{2}{\phi}}\lambda = u = e^{-\frac{i}{2}{\phi}}\Lambda, \quad
e^{-\frac{i}{2}{\phi}}\bl = \bu = e^{\frac{i}{2}{\phi}}\bL \, .
\ee
Now one may rewrite the constraints \p{ucon} in a symmetrical, linear (although disguisedly nonlinear) form
\bea\label{lcon}
D^i \lambda=0, \quad \bD_i \bl =0, &\qquad &\nabla^a \Lambda=0, \quad \bnab_a \bL=0, \nn
D^i \bL=-i\bD^i\phi, \; \bD_i \Lambda = -i D_i \phi, & \qquad & \nabla^a \bl=-i\bnab^a \phi, \; \bnab_a\lambda=-i\nabla_a\phi.
\eea

Now, it is time to analyze the component structure of our supermultiplet. First of all, let us define the
following components of our superfields:\footnote{The higher components in the superfields are easily expressed through
the time derivatives of those in \p{comp}.}

\bea
&&
\left. \lambda \right|_{\theta=0} = \lambda(t)\,, \quad
\left. \overline{\lambda} \right|_{\theta=0} = \overline{\lambda}(t)\,, \quad
\left. \phi \right|_{\theta=0} = \phi(t)\,,
\nn
&&
\left. D^i \overline{\Lambda} \right|_{\theta=0} = \xi^i(t)\,, \quad
\left. {\bD}_i \Lambda \right|_{\theta=0} = \bar \xi_i(t)\,, \quad
\left. \nabla^a \overline{\lambda} \right|_{\theta=0} = \psi^a(t)\,, \quad
\left. {\bnab}_a \lambda \right|_{\theta=0} = \bar \psi_a(t)\,, \nn
&&
\left. D^2 \overline{\Lambda} \right|_{\theta=0} = i\,B(t)\,, \quad
\left. {\bD}^2 \Lambda \right|_{\theta=0} = i\, \bar B(t)\,, \quad
\left. \nabla^2 \overline{\lambda} \right|_{\theta=0} = i\, A(t)\,, \quad
\left. {\bnab}^2 \lambda \right|_{\theta=0} = i\,\bar A(t)\,, \nn
&&
\left. D^i \nabla^a \overline{\lambda} \right|_{\theta=0}=Y^{ia}(t), \;
\left. \bD^i \bnab^a {\Lambda} \right|_{\theta=0}=-{\bar Y}^{ia}(t)\,.
\label{comp}
\eea
Not all of these components are independent. Indeed, one may check that the constraints \p{lcon} imply
the reality of $Y^{ia}$
\be
Y^{ia} = {\bar Y}^{ia}
\ee
and impose the following additional relations between the components $A$ and $B$:
\bea
&& A-{\bar A} = -4i{\dot\phi}, \quad B-{\bar B} = -4i{\dot\phi}, \label{dcon1} \\
&& \dot{A} + \dot{\bar B} =0. \label{dcon2}
\eea
Thus, one may see that among the four auxiliary components $A,B$ only one, namely
$(\mbox{Re } A -\mbox{Re } B)$, is
independent: the imaginary parts are expressed through $\dot{\phi}$ while the
real ones are  subjected to the differential constraint  following from \p{dcon2}
\be\label{dcon}
\frac{d}{dt} \left( \mbox{Re } A + \mbox{Re } B \right) =0 \; \Rightarrow \; \mbox{Re } A + \mbox{Re } B =m.
\ee
Therefore, our nonlinear supermultiplet contains eight bosonic components: three physical fields --
$\phi, \lambda,\bl$ and five auxiliary ones -- $Y^{ia}, (\mbox{Re } A - \mbox{Re } B) $, as well as
eight fermionic fields -- $\xi^i,{\bar\xi}{}^i, \psi^a,{\bar\psi}{}^a$.

Before going further to construct an invariant superfield action, let us point the attention on the two
essential features of our construction. First of all, we defined the components of our
nonlinear supermultiplet \p{comp} in a rather non standard way. Indeed, one may see that the
components are defined not only as the spinor derivatives of, say, superfields $\left\{ \phi, \lambda,\bl\right\}$,
but also through derivatives of another set of superfields  $\left\{ \phi, \Lambda,\bL\right\}$. Of course,
the latter ones are expressed through the former ones. However, proceeding with a such definition, the relations
between different auxiliary components in \p{comp} acquired the simplest form, as in \p{dcon1},\p{dcon2},\p{dcon}.
Secondly, one may see from \p{dcon} that the constant $m$ appears as a component of our supermultiplet.
As we already know from \cite{leva, bikl1,2dim}, the presence of this constant is crucial as for generating
the potential terms in the action, as well as for the dualization procedure \cite{dual1}.

While being a quite reasonable choice for establishing the
irreducible constraints for the superfields, the $N=8, d=1$
superspace is not too suitable for constructing the invariant
action. In one dimension, the $N=4$ superspace provides the best
framework for the superfield action. Fortunately enough, for the
case at hands the proper $N=4$ superspace is almost evident.
Indeed, analyzing the constraints \p{lcon}, one may note that
$\left\{ D^2,\bD_2, \nabla^2, \bnab_2 \right\}$ derivatives from
all superfields and all their combinations are expressed in terms
of $\left\{ D^2,\bD_2, \nabla^2, \bnab_2 \right\}$ derivatives
from another superfield. For example, the spinor components with
indices 2 defined in \p{comp} can be equivalently expressed as
follows: \be \xi^2 =-i\bD_1\phi,\quad \bar\xi_2=iD^1 \phi, \quad
\psi^2=i\bnab_1 \phi , \quad \bar\psi_2 =i\nabla^1 \phi\; . \ee
Thus, all independent components present in our $N=8$ nonlinear
superfields appear in the expansion over Grassmann variables with
index "1" only. Therefore, the proper action reads
\be\label{action1} S=\int dt d^2 \theta^1 d^2 \vartheta^1 {\cal L}
=\int dt D^1 \bD_1 \nabla^1 \bnab_1 {\cal L} \ee where an
arbitrary, for the time being, function ${\cal L}$ depends on
$\theta^2=\vartheta^2=0$ projections of our superfields $\phi,
u,\bu$ (or $\phi,\lambda,\bl$ or $\phi,\Lambda,\bL$). Of course,
by construction, the action \p{action1} is invariant only under
the manifest $N=4$ supersymmetry acting on $\theta^1,\vartheta^1$.
The invariance under implicit $N=4$ supersymmetry imposes the
following constraint on the function $L(\phi, u,\bu)={\cal
L}|_{\theta=\vartheta=0}$: \be\label{lapu} \frac{\partial^2
L}{\partial \phi^2}+\frac{1}{4}\left( u \frac{\partial L}{\partial
u}+\bu \frac{\partial L}{\partial \bu}\right) +\left( 1
+\frac{1}{2} u\bu\right) \frac{\partial^2 L}{\partial u \partial
\bu}+ \frac{1}{4} \left( u^2 \frac{\partial^2 L}{\partial
u^2}+\bu^2 \frac{\partial^2 L}{\partial \bu^2}\right) =0. \ee Of
course, it makes no differences which set of superfields is chosen
in the superfield Lagrangian density ${\cal L}$. From now on, let
us fix this dependence to be on the set $\left\{ \phi,
\lambda,\bl\right\}$. With this choice the constraint \p{lapu}
acquires the standard form of the three dimensional Laplace
equation on the sphere $S^3$ in stereographic coordinates \be
\Delta_3 L = (1+ \lambda \overline{\lambda})\;\frac{\partial^2
L}{\partial \lambda \partial \overline{\lambda}} +
\frac{\partial^2 L}{\partial \phi^2} + i
\overline{\lambda}\;\frac{\partial^2 L}{\partial
\overline{\lambda} \partial \phi} -  i \lambda\;\frac{\partial^2
L}{\partial \lambda \partial \phi} = 0\,. \label{lapl} \ee Thus we
conclude that the action \p{action1} with the function ${\cal L}
(\phi, \lambda, \bl)$ obeying \p{lapl} is invariant under the
entire $N=8$ supersymmetry.

For completing this Section, let us note that the  constraints on the superfields $\left\{ \Lambda,\bL \right\}$
in \p{lcon} may be written as
\be
D^i \Lambda +\Lambda \bD{}^i \Lambda =0, \quad \nabla^a \Lambda =0.
\ee
The first part of these constraints coincides with the constraints on the
$N=4$ nonlinear chiral supermultiplet \cite{2dim,8nlin,2dimc}, while the second part is just a
chirality conditions. So, one may wonder whether it is possible to write the superfield action for our
supermultiplet in a way similar to the action for the nonlinear chiral supermultiplet,
as an integral over the chiral superspace
\be\label{action2}
S_{chir} \sim \int dt d^2 \theta d^2 \vartheta \cbF(\bL) + \int dt d^2 \bar\theta d^2 \bar \vartheta \cF(\Lambda),
\ee
where $F(\Lambda)$ is an arbitrary function depending on the superfield $\Lambda$ only. The simplest consideration shows that the
action \p{action2} is perfectly invariant with respect to full $N=8$ supersymmetry.
The natural question is how the action \p{action2}
is related with the action \p{action1} which is supposed to be the most general one. In order
to clarify this point, let us consider an arbitrary term in the expansion of integrand in \p{action2}
over $\Lambda$ and perform the following transformations:
\bea
\bnab^2 \bD^2\left( \frac{a_n}{n!} \Lambda^n\right)& \sim & \bnab^2 \bD_1 \bD_2 \left( \frac{a_n}{n!} \Lambda^n\right) =
\bnab^2 \bD_1  \left( \frac{a_n}{(n-1)!} \Lambda^{n-1} \bD_2 \Lambda \right) =
 \bnab^2 \bD_1\left( -i\frac{a_n}{(n-1)!} \Lambda^{n-1}D^1 \phi\right) \nn
&& = \bnab^2 \bD_1
 \left( -i\frac{a_n}{(n-1)!} \lambda^{n-1} e^{i(n-1)\phi} D^1 \phi\right) =\bnab^2 \bD_1 D^1 \left( -\frac{a_n}{(n-1)(n-1)!} \Lambda^{n-1}\right).
\eea
Playing a similar game with $\bnab_2$ derivatives we will finally transform the integrand as follows:
\be
\bnab^2 \bD^2\left( \frac{a_n}{n!} \Lambda^n\right) \sim \bnab_1 \nabla^1 \bD_1 D^1 \left( e^{i\phi} \left( 1+\Lambda \bL \right) \Lambda^{n-2}\right).
\ee
Therefore, the full action \p{action2} can be rewritten in the form given in \p{action1} with the Lagrangian
\be\label{lchir}
{\cal L}_{chir} = \frac{1+\Lambda\bL}{\Lambda\bL} \left( e^{i\phi} \cF' \bL + e^{-i\phi} \cbF' \Lambda \right).
\ee
It is a matter of straightforward calculations to check that \p{lchir} obeys the Laplace equation
(in the variables $\phi,\Lambda,\bL$)
\be
\Delta_3 L = (1+ \Lambda \overline{\Lambda})\;\frac{\partial^2 L}{\partial \Lambda \partial \overline{\Lambda}}
+ \frac{\partial^2 L}{\partial \phi^2}
+ i \overline{\Lambda}\;\frac{\partial^2 L}{\partial \overline{\Lambda} \partial \phi}
-  i \Lambda\;\frac{\partial^2 L}{\partial \Lambda \partial \phi} = 0\,,
\label{laplll}
\ee
as it should be. Thus, the action \p{action2} is a particular case of the more general action \p{action1}.
\setcounter{equation}0
\section{Bosonic sector, potential terms and dualization}

The full component action is rather lengthy, but its pure bosonic core is remarkably simple. Integrating in \p{action1}
over Grassmann variables, discarding all fermions and eliminating the auxiliary fields $Y^{ia}, {\mbox Im }A, {\mbox Im }B$
by means of their equations of motion, we get the following action:
\bea\label{action_comp1}
S& = & \int dt \frac{\partial^2 L}{\partial \lambda \partial \bl}\left[ (\mbox{Re }A) \partial_t \left( \lambda\bl\right) +
4 \dot\lambda \dot\bl + 2i \dot{\phi} \left( \lambda \dot\bl- \dot\lambda \bl\right)+\left(1+\lambda\bl\right)\left(\frac{1}{4}
(\mbox{Re }A)(\mbox{Re }B) +\dot\phi{}^2\right)  \right] \nonumber \\
 & & +  (\mbox{Re }A + \mbox{Re }B) \int dt \left[ \frac{i}{2} \left( \frac{\partial^2 L}{\partial \phi \partial \lambda} \dot\lambda -
\frac{\partial^2 L}{\partial \phi \partial \bl} \dot\bl \right) -\frac{1}{2} \dot\phi \left(
\frac{\partial^2 L}{\partial \phi \partial \lambda} \lambda +
\frac{\partial^2 L}{\partial \phi \partial \bl} \bl \right) \right].
\eea
Before imposing the last constraint \p{dcon} and eliminating the last auxiliary component, let us introduce the new function
$h(\phi, \lambda,\bl)$ as follows:
\be\label{h}
h(\phi, \lambda,\bl) = -\frac{\partial L}{\partial \phi} + i\left( \lambda \frac{\partial L}{\partial \lambda}-\bl \frac{\partial L}{\partial \bl}\right).
\ee
One may check that this function $h$ also obeys the three dimensional Laplace equation on the sphere $S^3$
\be\label{lap_h}
(1+\lambda\bl) h_{\lambda\bl}+h_{\phi\phi}-i\lambda h_{\phi\lambda}+i\bl h_{\phi\bl} =0.
\ee
Now, rewriting the action \p{action_comp1} in terms of $h$, taking into account the constraint \p{dcon} and eliminating the last
auxiliary component, we end up with the bosonic action
\be\label{action_comp2}
S  =  \int dt  \left\{ h_\phi\left[\frac{4 \dot\lambda \dot\bl}{(1+\lambda\bl)^2}+ \left( \dot\phi +i \frac{\lambda\dot\bl -\dot\lambda \bl}
{1+\lambda\bl}\right)^2 -\frac{1}{16} m^2  +
\frac{m}{2}  \frac{\partial_t\left( \lambda\bl\right)}{1+\lambda\bl}\right]+i\frac{m}{2}\left( h_\bl \dot\bl -h_\lambda \dot\lambda\right)
\right\}.
\ee
The bosonic kinetic terms of the action \p{action_comp2} describe just the sphere $S^3$ in
stereographic coordinates modified by an arbitrary function $h_\phi$, which is a harmonic function on $S^3$.
In addition, the action contains potential terms which are completely specified by the
same function $h$. One should stress that the action \p{action_comp2} is very similar to the bosonic part of
the action describing the $N=4$ supersymmetric particle on the sphere $S^3$ \cite{nlin,nlin1,nlin2}. The
essential difference is that the metrics and the potential terms are defined by the {\it same} harmonic
function $h$, while in the $N=4$ supersymmetric case the metrics and potential terms are not related. Thus,
$N=8$ supersymmetry puts a rather strong restrictions on the possible potential terms. As an impressive
example, one may consider the particle on the $S^3$ itself. This case corresponds to $h_\phi=1$,
and one may immediately check that all potential terms disappear, being either constants or full time
derivatives. This is just the result we got in Section 2, while performing the reduction from quaternionic
$N=8$ supersymmetric mechanics.

The last issue we are going to discuss in this Section is a dualization of the coupling constant $m$ entering
the action \p{action_comp2} into a fourth physical bosonic field. Following \cite{dual2}, we treat the
self-evident statement $m=const$ as the additional constraint $\partial_t m =0$. If we include this
constraint into our action \p{action_comp2} with a Lagrangian multiplier,
it will be possible to express the ``former constant'' $m$ in terms of the Lagrange multiplier. So, we have
\be\label{act11}
{\hat S}=S +\frac{1}{2} \int\; dt\;  m\;  \dot{u} \;\quad \Rightarrow  \quad
-\frac{1}{8}h_\phi m  +
\frac{1}{2}  \frac{h_\phi\partial_t\left( \lambda\bl\right)}{1+\lambda\bl}+i\frac{1}{2}\left( h_\bl \dot\bl -h_\lambda \dot\lambda\right)
+ \frac{1}{2}\dot u=0 .
\ee
Now we  plug this expression for $m$ back into the action ${\hat S}$. Doing in a such way
we get a four dimensional sigma-model action with the following bosonic part:
\be\label{finS}
{\hat S}=\int dt\; \left\{ h_\phi\left[ \frac{4 \dot\lambda\dot\bl}{(1+\lambda\bl)^2} +
\left( \dot\phi +i \frac{ \dot\bl \lambda - \dot\lambda \bl }{1+\lambda\bl}\right)^2\right] +
\frac{1}{h_\phi}\left[ {\dot u} +i \left( \dot\bl h_\bl -\dot\lambda h_\lambda\right)\right]^2 \right\}.
\ee
The action \p{finS} depends on one arbitrary function  $h(\phi, \lambda,\bl)$,
which obeys the Laplace equation on $S^3$. It is interesting to note that the action \p{finS} exhibits the
same target space geometry   which appears in the heterotic $(4,0)$ sigma-model
in $d=2$ \cite{heterotic}.

In the particular case with $h_\phi=1, h_\lambda=0$ the action \p{finS} is reduced to
the direct sum of the free  actions on $S^3$ and on $S^1$. Another interesting limit
corresponds to a linearized version of the action \p{finS}. In this case it acquires the form
of the Gibbons-Hawking Ansatz for the hyper-K\"ahler sigma model action \cite{GH}:
\be\label{finS1}
{\tilde S}=\int dt\; \left\{ h_\phi\left[ 4 \dot\lambda\dot\bl +
\left( \dot\phi +i \left( \dot\bl \lambda - \dot\lambda \bl \right) \right)^2\right] +
\frac{1}{h_\phi}\left[ {\dot u} +i \left( \dot\bl h_\bl -\dot\lambda h_\lambda\right)\right]^2 \right\},
\ee
with the function $h$ obeying the ``flat'' three dimensional Laplace equation
\be\label{lap_h1}
h_{\lambda\bl}+h_{\phi\phi}=0.
\ee

\section{Conclusions}

In this paper, proceeding from quaternionic $N=8$ supersymmetric mechanics, we performed a reduction over a
bosonic radial variable, ending up with a nonlinear off-shell supermultiplet with three bosonic end eight
fermionic physical degrees of freedom. The simplest action describes the $N=8$ supersymmetric mechanics with
the sphere $S^3$ in the bosonic part of the action. We find the irreducible constraints on the supermultiplet
arising upon this reduction and construct the most general superfield action for three dimensional
supersymmetric mechanics. The geometry of the bosonic sector of the most general sigma-model type action is
described by an arbitrary function obeying the three dimensional Laplace equation on the sphere $S^3$.
Among the bosonic components of this new supermultiplet there is a constant which gives rise to potential terms.
After dualization of this constant one may come back to the supermultiplet with four physical bosons.
However, this new supermultiplet is highly nonlinear.

An obvious project for future study is to investigate the same
type of ``radii'' reductions for another linear $N=8$
supermultiplets \cite{bikl2, ios}. One may expect to find new
nonlinear supermultiplets which will exhibit a new type of
geometry in the bosonic sector. Another related interesting
question concerns the superfield description of the
supermultiplets depending on the arbitrary function, like our four
dimensional nonlinear supermultiplet, which we constructed in the
present paper by dualization of the coupling constant. Finally, it
would be rather interesting to check the integrability of the
constructed system for the simplest choices of the harmonic
function $h_\phi$ in a full analogy with the results presented in
\cite{GW}.

\section*{Acknowledgements}

S.K. and A.S. thank the INFN-Laboratori Nazionali di Frascati,
where this work was completed, for warm hospitality. This work was
partly supported by grants RFBR-06-02-16684, 06-01-00627-a, DFG
436 Rus 113/669/03 and by INTAS under contract 05--7928.

\end{document}